\documentclass[twocolumn,showpacs,prb]{revtex4-1}
\usepackage{bm}
\usepackage{graphicx}
\usepackage{epsfig}
\usepackage{amsmath} 
\usepackage{color}
\usepackage{ulem}
\usepackage{amssymb}
\usepackage[colorlinks = true]{hyperref}

\renewcommand{\i}{\mathrm{i}}
\newcommand{\eps}{\varepsilon}
\newcommand{\e}{\mathrm{e}}

\newcommand{\bracket}[3]{\left \langle #1 \left| #2 \right| #3 \right \rangle}

\newcommand{\B}{\mathcal B}
\newcommand{\D}{\mathcal D}
\newcommand{\A}{\mathcal A}

\definecolor{darkgray}{rgb}{0.66, 0.66, 0.66}

\begin{document}

\title{Optical properties of helical edge channels in zinc-blende-type topological insulators:
Selection rules, circular and linear dichroism, circular and linear photocurrents}

\author{M.\,V.\,Durnev}
\author{S.\,A.\,Tarasenko}

\affiliation{Ioffe Institute, 194021 St.\,Petersburg, Russia}

\begin{abstract}
We develop a theory of electron-photon interaction for helical edge channels in two-dimensional topological insulators based on zinc-blende-type
quantum wells. It is shown that the lack of space inversion symmetry in such structures enables the electro-dipole optical transitions between the 
spin branches of the topological edge states. Further, we demonstrate the linear and circular dichroism associated with the edge states and
the generation of edge photocurrents controlled by radiation polarization.
\end{abstract}

\maketitle 

\section{Introduction}

The study of conducting edge channels with spin-momentum locking which are inherent to two-dimensional electron systems with non-trivial topology is one of the central topics in the physics of topological insulators (TIs)~\cite{Bernevig2006,Konig2007,Knez2011}. Much effort is being invested now into the study of transport properties of edge channels such as local and non-local conductivity~\cite{Roth2009,Gusev2011,Ma2015,Tikhonov2015}, injection of carriers from edge states into magnetic materials or superconductors~\cite{Hart2014,Kononov2015}, and the mechanisms of 
backscattering~\cite{Tanaka2011,Lunde2012,Altshuler2013,Vayrynen2014,Entin2015,Kurilovich2017}.
Optical studies of helical edge channels, although being challenging, are also in high demand since they can provide insight into the spin structure of the edge states and details of electron-photon interaction. It was experimentally demonstrated recently that the photoionization of edge channels by polarized terahertz radiation is asymmetric in $\bm k$-space and is accompanied by the emergence of edge photocurrents~\cite{Dantscher2017}. It was also proposed theoretically that radiation with the photon energy smaller than the bulk gap can induce direct optical transitions between the ``spin-up'' and ``spin-down'' branches of the helical channel and excite a photocurrent circulating around the sample edges~\cite{Dora2012, Artemenko2013}. Previous research of the inter-branch optical transitions was phenomenological and based on a centro-symmetric model of TIs which allows only (weak) magneto-dipole coupling of
the ``spin-up'' and ``spin-down'' states by the magnetic field of the radiation~\cite{Dora2012, Artemenko2013}. However, the practical realization of two-dimensional TIs is II-VI (HgTe/CdHgTe) or III-V (InAs/GaSb) zinc-blend-type structures with the natural lack of the space inversion center in the crystal lattice~\cite{Konig2007,Knez2011}. In particular, in the most studied TIs based on HgTe/CdHgTe quantum wells 
(QWs), the strong natural interface inversion asymmetry leads to the mixing of the ``spin-up'' and ``spin-down'' states at the QW interfaces~\cite{Tarasenko2015}. The mixing considerably modifies the energy spectrum of ''bulk'' states as well as the structure and magnetic properties of helical edge channels~\cite{Durnev2016}.

\begin{figure}[t]
\includegraphics[width=0.95\linewidth]{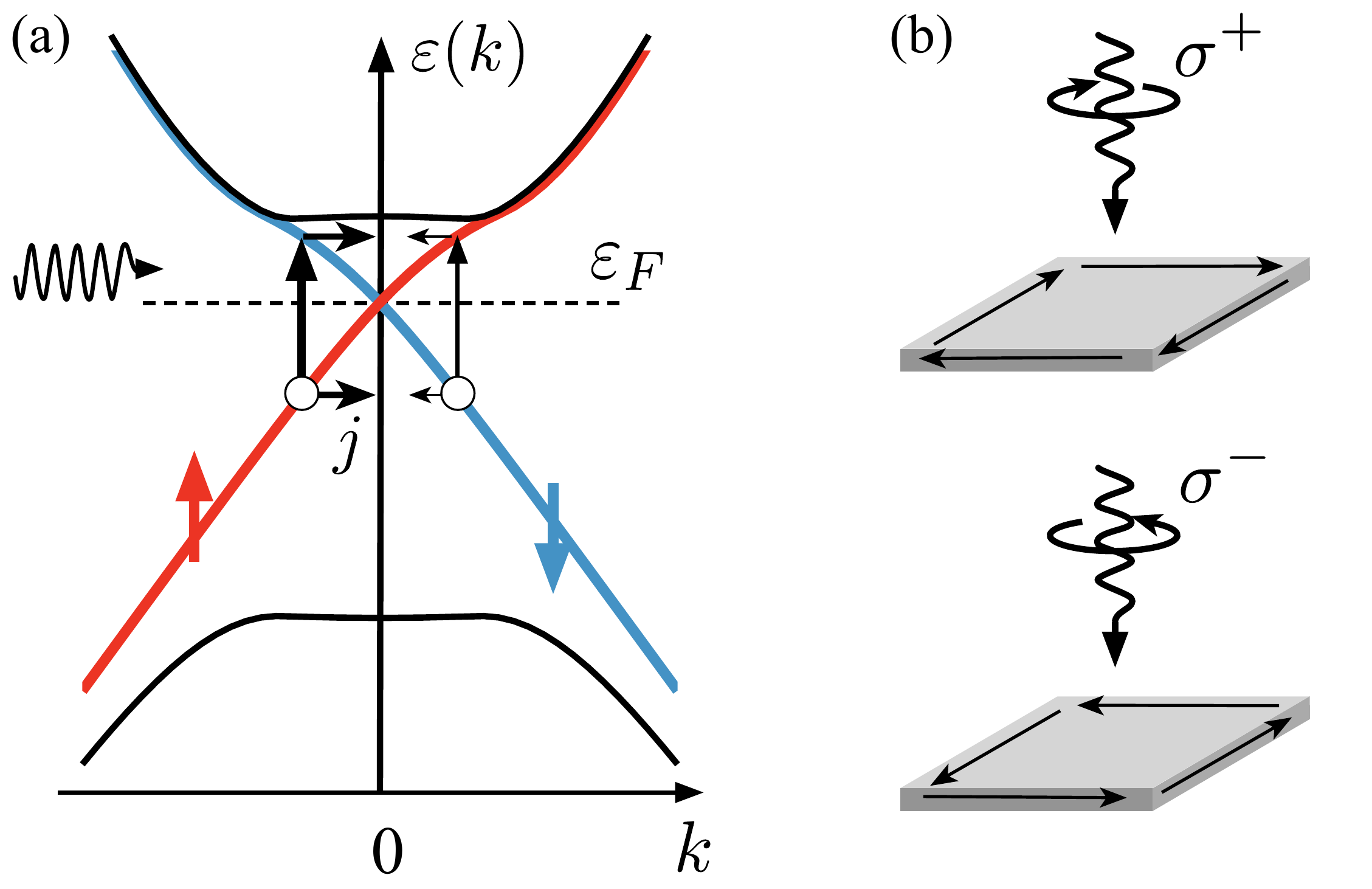}
\caption{\label{fig:fig2} 
(a) Sketch of electron dispersion in a two-dimensional topological insulator. The ``spin-up'' and ``spin-down'' branches 
of the edge-state dispersion are shown by red and blue curves, respectively, $k$ is the wave vector along the edge. Optical transitions
between the spin branches induced by polarized radiation occur at $k$ and $-k$ at different rates, which leads to a direct electric current
in the edge channel. (b) Edge photocurrents excited by normally incident circularly polarized radiation flow in the opposite directions
for the right-handed and left-handed polarizations.
}
\end{figure}

Here, we describe the optical properties of helical edge channels in zinc-blend-type two-dimensional TIs.
We show that, in such systems, direct optical transitions between the ``spin-up'' and ``spin-down'' branches of the edge-state dispersion occur not only in the magneto-dipole approximation but also in the much stronger electro-dipole mechanism of the electron-photon-interaction. Moreover, the probability of the absorption of circularly polarized photons is asymmetric in $\bm k$-space which leads to the circular photogalvanic effect where the transfer of the photon angular momenta to the electrons drives a direct electric current, see Fig.~\ref{fig:fig2}.  The interference of the electro-dipole and the magneto-dipole mechanisms of the photon absorption for circularly polarized radiation is constructive or destructive depending on the photon helicity, which leads to the circular dichroism. For linearly polarized radiation, the interference gives rise to an asymmetry of the optical transitions in $\bm k$-space and to a linear photocurrent. In this case, the photocurrent direction depends on the edge crystallographic orientation and the radiation polarization vector.

The paper is organized as follows. In Sec.~\ref{Sec2}, we present a symmetry consideration and develop a microscopic theory of the electron-photon interaction for helical edge states in zinc-blend-type TIs. In the framework of the extended Bernevig-Hughes-Zhang model with the interface terms included we calculate the matrix elements of the electric dipole and magnetic dipole operators for HgTe/CdHgTe-based TIs. In Sec.~\ref{Sec3}, we consider the optical transitions between 
the ``spin-up'' and ``spin-down'' branches of the edge-state dispersion and develop a theory of the linear and circular dicroisms. Section~\ref{Sec4} is devoted to the theory of
the edge photocurrents excited by circularly and linearly polarized radiation.

\section{Electron-photon interaction in helical channels}\label{Sec2}

\subsection{Edge states. General symmetry analysis}\label{edge-symmetry}

We consider a two-dimensional topological insulator based on zinc-blende-type QW.
The structure supports a pair of conducting helical edge states in the (topologically non-trivial) gap of the QW, see Fig.~\ref{fig:fig2}. The edge states are characterized by the wave vector $k_y$ directed along the edge and the pseudospin index $s = \pm 1/2$ enumerating the branches. We use the coordinate frame ($xyz$), where the $x$ axis is in the QW plane, perpendicular to the edge and is pointing inside the sample, $y$ is parallel to the edge, and $z$ is the QW growth axis. At small $k_y$, the dispersion of the edge states is linear: $\eps_{k_y \pm 1/2} = \pm \hbar v_0 k_y$, where $v_0$ is the velocity.  The states $|k_y , s\rangle$ and $|-k_y , -s\rangle$ are related by time reversal symmetry and, therefore, have the same energy (Kramers degeneracy). We take the corresponding wave functions $\psi_{k_y s}$ to satisfy the relation 
\begin{equation}\label{T-symmetry}
{\cal T} \psi_{k_y s} = - 2s  \, \psi_{-k_y -s} \,,
\end{equation}
where ${\cal T}$ is the operator of time reversal. The operator ${\cal T}$ commutes with the Hamiltonian, satisfies ${\cal T}{\cal T} = -1$, and can be presented in the form ${\cal T} = U_{t} K$, where $U_{t}$ is a unitary operator, i.e., $U_{t}^\dag = U_{t}^{-1}$, and $K$ is the operator of complex conjugation.

\begin{figure}[htpb]
\includegraphics[width=0.47\textwidth]{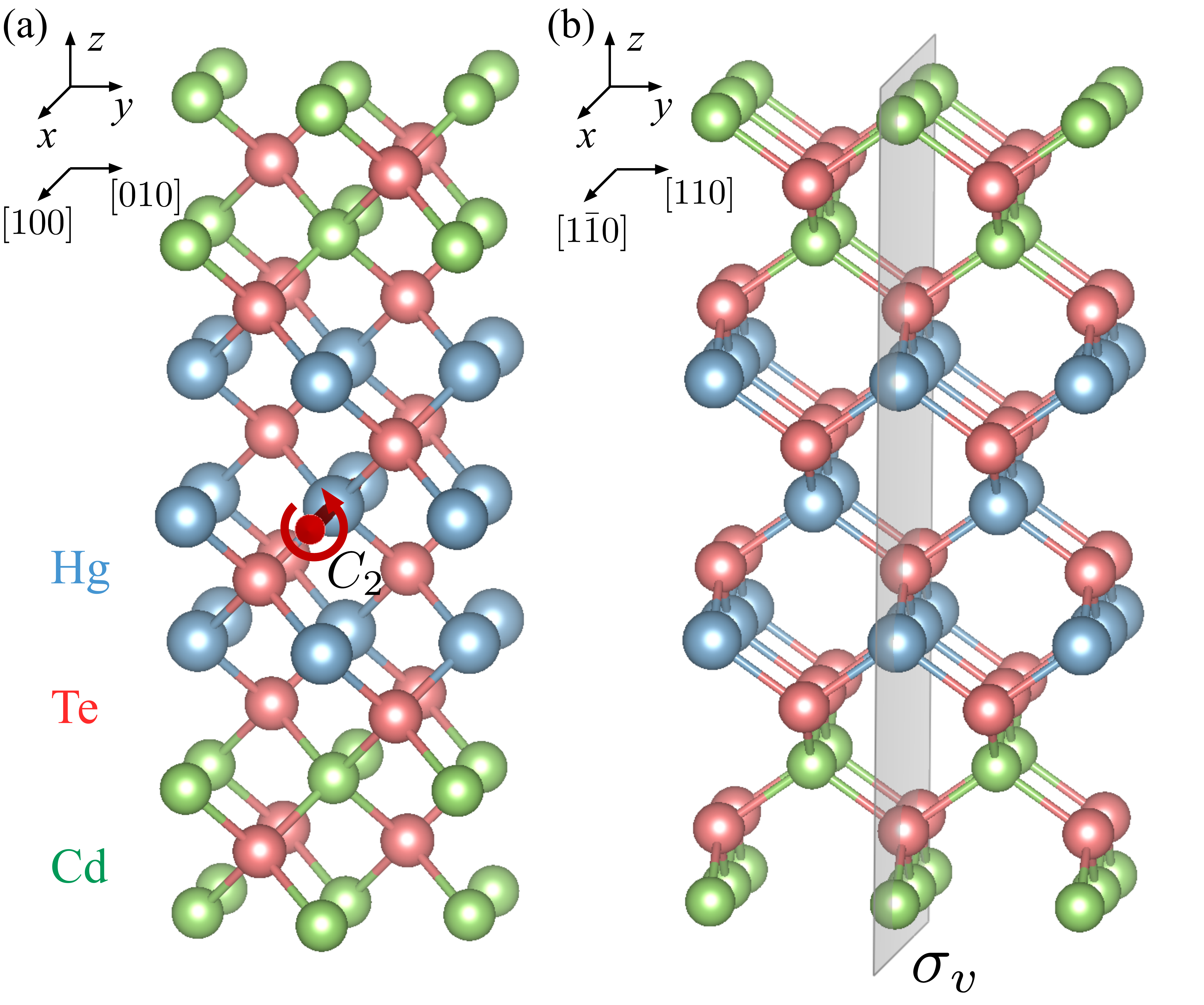}
\caption{\label{fig:fig1}
Side view of HgTe/CdTe quantum wells with the edge along (a) [010] and (b) [110] axes. The structures are described by the $C_2$ and $C_s$ point groups, respectively. The corresponding symmetry elements, the rotation axis $C_2$ (a) and the mirror plane $\sigma_v$ (b), are shown by the red arrow and the gray plane.
}
\end{figure}

Additional information about the edge states can be obtained from the spatial symmetry of the structure. The point-group symmetry
of an infinite (001)-grown QW with a symmetric heteropotential is $D_{2d}$. This point group takes into account the lack of a space inversion center in the QW due to the bulk inversion asymmetry of the host crystal and the inversion asymmetry at QW interfaces~\cite{Durnev2016}. 
Introduction of an edge lowers the spatial symmetry of the system. For an arbitrary orientation of the edge with respect to crystallographic axes, the point-group symmetry reduces to the trivial group $C_1$ with no non-trivial symmetry elements.
However, for two particular classes of the structures with the edges directed along $\langle$100$\rangle$ and $\langle$110$\rangle$ axes, which are commonly studied, the point-group symmetry is higher and contains non-trivial elements. Figure~\ref{fig:fig1} illustrates the crystal structures of (001)-grown HgTe/CdTe QWs with the edges along these high-symmetry directions.

The QW structure with the edge parallel to [010] (or similar direction) is described by the $C_2$ point group with the two-fold rotation axis $C_2 \parallel [100]$ at the center of the QW, see Fig.~\ref{fig:fig1}a. The corresponding operator of rotation by the angle $\pi$ about the $x$ axis ${\cal R}$  commutes with the Hamiltonian and relates the states $|k_y , s\rangle$ and $|-k_y , -s\rangle$. The rotation operator also satisfies ${\cal R}{\cal R} = -1$ and ${\cal R}^\dag = {\cal R}^{-1}$. By a proper choice of the phase of the wave functions we made them to satisfy the relation  
\begin{equation}\label{R-rotation}
{\cal R} \psi_{k_y s} = -i  \psi_{-k_y -s} \,
\end{equation}
and, in particular, ${\cal R} \psi_{0 \pm 1/2} = -i  \psi_{0 \mp 1/2}$. The latter relation corresponds to the usual transformation rule of spinors under the rotation by $\pi$ about the $x$ axis. It follows that the Pauli matrix $\sigma_x$ acting in the pseudo-spin space with the basis functions $\psi_{0 +1/2}$ and $\psi_{0 -1/2}$ is invariant under the rotation (belongs to the $\Gamma_1$ irreducible representation of the $C_2$ group) whereas the the Pauli matrices $\sigma_y$ and $\sigma_z$ change their sign under the rotation (belong to the $\Gamma_2$ irreducible representation)~\cite{koster63}.
All components of the polar and axial vectors in the $C_2$ group also transform according to either $\Gamma_1$ or $\Gamma_2$ representations, which is summarized in Tab.~\ref{tab1}.

The (001)-grown QW structure (also with an asymmetric confinement potential) with the edge parallel to [110] is described by the $C_{s}$ point group which contains the mirror plane $\sigma_v \parallel (110)$, see Fig.~\ref{fig:fig1}b. The operator of reflection in the $(xz)$
plane $\tilde{{\cal R}}$ commutes with the Hamiltonian and satisfies $\tilde{{\cal R}}\tilde{{\cal R}} = 1$ and $\tilde{{\cal R}}^\dag = \tilde{{\cal R}}^{-1}$. The wave functions $\psi_{k_y s}$ can be chosen in the way that
\begin{equation}\label{R-reflection}
\tilde{{\cal R}} \psi_{k_y s} = 2 i s \, \psi_{-k_y -s} \,.
\end{equation}
This equation at $k_y =0$ corresponds to the usual transformation rule of spinors under the reflection in the $(xz)$
plane. In this case, the Pauli matrix $\sigma_y$ acting in the pseudo-spin space with the basis functions $\psi_{0 +1/2}$ and $\psi_{0 -1/2}$ transforms according to the $\Gamma_1$ irreducible representation of the $C_s$ group whereas the Pauli matrices $\sigma_x$ and $\sigma_z$ transform according to the $\Gamma_2$ irreducible representation~\cite{koster63}. The classification of the polar and axial vector components according to the representations they transform by is summarized in Tab.~\ref{tab1}.

\begin{table}[htb]
\caption{\label{tab1} Irreducible representations and basis functions constructed from the components of the polar $\bm r = (x,y,z)$ and axial $\bm B = (B_x, B_y, B_z)$ vectors and the Pauli matrices for the $C_2$ and $C_s$ point groups.
}
\begin{tabular}{c|c|c}
\hline
\hline
rep & basis functions & basis functions \\
 & C$_2$ ($x \parallel [100]$, $y \parallel [010]$) & C$_s$ ($x \parallel [1\bar{1}0]$, $y \parallel [1 1 0]$) \\
 \hline
$\Gamma_1$ & $x$, $B_x$, $\sigma_x$ & $x$, $z$, $B_y$, $\sigma_y$ \\
$\Gamma_2$ & $y$, $z$, $B_y$, $B_z$, $\sigma_y$, $\sigma_z$ & $y$, $B_x$, $B_z$, $\sigma_x$, $\sigma_z$\\
\hline
\hline
\end{tabular}
\end{table}

Let us now construct the effective Hamiltonian of edge states at small $k_y$ using the method of invariants~\cite{birpikus}. From the requirement that the Hamiltonian is invariant with respect to all symmetry operations and time reversion follows that, to the first order in $k_y$, the Hamiltonian can contain terms $\sigma_z k_y$ and $\sigma_y k_y$ in the structure of the $C_2$ point group and the terms  
 $\sigma_z k_y$ and $\sigma_x k_y$ in the structure of the $C_s$ point group. However, by a unitary transformation we can convert
the effective Hamiltonian to the form $\propto \sigma_z k_y$ removing other $k_y$-linear terms. Indeed, the basis $(\psi_{0 +1/2}, \psi_{0 -1/2})$ is not fully determined yet. Any pair of the functions of the form 
$(\alpha \psi_{0 +1/2} + \beta \psi_{0 -1/2}, \, \alpha^* \psi_{0 -1/2} - \beta^* \psi_{0 +1/2})$ also satisfies Eq.~\eqref{T-symmetry} together with Eq.~\eqref{R-rotation} [or together with Eq.~\eqref{R-reflection}] provided $|\alpha|^2+|\beta|^2 = 1$ and $\alpha$ is real and $\beta$ is imaginary (or both $\alpha$ and $\beta$ are real). By a proper choice of the wave function basis we can exclude the term $\sigma_y k_y$ (or $\sigma_x k_y$) from the Hamiltonian. 
Then, the effective Hamiltonian of edge states up to the third order in $k_y$ reads
\begin{equation}
\label{eq:H0}
\mathcal H_{\rm edge} =  a \sigma_z k_y + b k_y^2  + \sum_{\alpha=x,y,z} c_{\alpha} \sigma_{\alpha} k_y^3 \:,
\end{equation}
where $a =\hbar v_0$, $b$ and $c_{\alpha}$ are real parameters; $c_x$ ($c_y$) is zero in the structure of the $C_2$ ($C_s$) point group.

\subsection{Electro-dipole interaction}

The Hamiltonian of electron-photon interaction can be also constructed using the method of invariants. In the electro-dipole approximation, the Hamiltonian of electron-photon interaction has the form 
\begin{equation}\label{eq:Ham_ed}
\mathcal H_{\rm edge}^{(\bm E)} = -\bm d \cdot \bm E \,,
\end{equation}
where $\bm E$ is the electric field of the electromagnetic wave and $\bm d$ is the electric dipole operator. Using Tab.~\ref{tab1}
and the time reversal symmetry one can construct $\bm d$ in the ($\psi_{0 +1/2}, \psi_{0 -1/2}$) basis. At $k_y=0$, the time reversal symmetry does not allow any terms coupling
the states with $s =\pm 1/2$. At $k_y \neq 0$, such a coupling is possible and the vector $\bm d$ to the first order in $k_y$ has the form
\begin{equation}\label{d_C2}
\bm d = \left(
\begin{array}{c}
D_{xy} \sigma_y + D_{xz} \sigma_z \\ D_{yx} \sigma_x  \\ D_{zx} \sigma_x 
\end{array}
\right) \times k_y
\end{equation}
in the structure of the $C_2$ group and the form
\begin{equation}\label{d_Cs}
\bm d = \left(
\begin{array}{c}
D_{xx} \sigma_x + D_{xz} \sigma_z \\ D_{yy} \sigma_y  \\ D_{zx} \sigma_x + D_{zz} \sigma_z 
\end{array}
\right) \times k_y
\end{equation}
in the structure of the $C_s$ group. Here, $D_{\alpha\beta}$ are real linearly independent parameters.
In a structure with an arbitrarily oriented edge, there are no symmetry restrictions and any component of the vector $\bm d$ can contain terms with any Pauli matrix. The terms proportional to $\sigma_z k_y$ describe the electric-field-induced change in the edge-state velocity while other terms lead to optical transitions between the spin branches. 

We emphasize that the electro-dipole optical transitions induced by the radiation polarized in the interface plane are possible only due to the low 
symmetry of the QW. Indeed, in the model of a QW with symmetric confinement potential and made of isotropic material~\cite{Teppe2018}, 
the point-group symmetry of this QW structure with an edge would be $C_{2v}$, which has both vertical and horizontal mirror planes and a two-fold rotation axis. In such a system, there are no terms in the dipole components $d_x$ and $d_y$ that couple the edge states with $\pm 1/2$ and, therefore, no optical transitions by the in-plane-polarized radiation occur in the electro-dipole mechanism. Note, that the same $C_{2v}$ point group describes (110)-oriented symmetrically-grown zinc-blende-type QW structures with the edge parallel to $[1\bar{1}0]$. Thus, the electro-dipole transitions are forbidden by symmetry in such structures as well.

The probability of the direct optical transitions between the edge states $\psi_{k_y s}$ and $\psi_{k_y -s}$ in the electro-dipole approximation is determined by the matrix elements of the operator $\bm d$. Instead of calculating the matrix elements of $\bm d$ it is more convenient sometimes to calculate the matrix elements of the velocity operator $\bm v$ which are related by $\bm v_{s\, -s} = \i (\omega_{s \, -s}/e) \bm d_{s \, -s}$, where $\omega_{s \, -s} = (\eps_{k_y s} - \eps_{k_y -s})/\hbar$ and $e$ is the electron charge. It follows from Eqs.~\eqref{d_C2} and~\eqref{d_Cs} that, at small $k_y$, the inter-branch matrix elements of the velocity operator are quadratic in $k_y$ and given by
\begin{equation}\label{v_C2}
\bm v_{s \, -s} =   \left(
\begin{array}{c}
D_{xy} \\ 2\i s D_{yx}  \\ 2 \i s D_{zx}  
\end{array}
\right) \times \frac{2 v_0 k_y^2}{e} 
\end{equation}
in the structure of the $C_2$ group and
\begin{equation}\label{v_Cs}
\bm v_{s \, -s} =   \left(
\begin{array}{c}
2\i s D_{xx} \\ D_{yy}  \\ 2 \i s D_{zx}  
\end{array}
\right) \times \frac{2 v_0 k_y^2}{e} 
\end{equation}
in the structure of the $C_s$ group.

The matrix elements~\eqref{v_C2} and~\eqref{v_Cs} satisfy the general relation
\begin{equation}
\label{Tsym}
\bm v_{s \, -s} (k_y) = \bm v_{-s \, s}^*(-k_y) \:
\end{equation}
which is imposed by time reversal symmetry and valid for edges of any orientation and arbitrary $k_y$. This relation follows from Eq.~\eqref{T-symmetry} and ${\cal T} \bm v = - \bm v {\cal T}$ which yield $\langle \psi_{k_y s} |\bm v | \psi_{k_y s'} \rangle =\langle 2s {\cal T} \psi_{-k_y -s} |\bm v | 2s' {\cal T}\psi_{-k_y -s'} \rangle = - 4 ss' \langle {\cal T}\psi_{-k_y -s} | {\cal T} \bm v \psi_{-k_y -s'} \rangle = - 4ss' \langle \psi_{-k_y -s} | \bm v | \psi_{-k_y -s'} \rangle^*$. Since $\bm v$ is an Hermitian operator, Eq.~\eqref{Tsym} yields $\bm v_{s \, -s} (k_y) = \bm v_{s \, -s}(-k_y)$, i.e., the inter-branch  matrix elements of the velocity operator are even in $k_y$.

The matrix elements of the velocity operator in the $C_2$ group, Eq.~\eqref{v_C2}, additionally satisfy the relations $v_{s \, -s}^{(x)} (k_y) = v_{-s \, s}^{(x)} (-k_y)$ and $v_{s \, -s}^{(y,z)} (k_y) = - v_{-s \, s}^{(y,z)}(-k_y)$
 imposed by the two-fold rotation axis, see Eq.~\eqref{R-rotation}. Combining them with Eq.~\eqref{Tsym} we obtain that in the $C_2$ group
\begin{equation}
\label{vC2_props}
v_{s \, -s}^{(x)} {\rm \; are \; real} \:,\:\: v_{s \, -s}^{(y,z)} {\rm \; are \; imaginary} \:.
\end{equation}

Similarly, the matrix elements of the velocity operator  in the $C_s$ group, Eq.~\eqref{v_Cs}, satisfy the relations 
$v_{s \, -s}^{(x,z)}(k_y) = - v_{-s \, s}^{(x,z)} (-k_y)$ and $v_{s \, -s}^{(y)} (k_y) = v_{-s \, s}^{(y)} (-k_y)$ imposed
by the mirror plane, see Eq.~\eqref{R-reflection}. Combining them with Eq.~\eqref{Tsym} we conclude that in the $C_s$ group
\begin{equation}
\label{vCs_props}
v_{s \, -s}^{(x,z)} {\rm \; are \; imaginary} \:,\:\: v_{s \, -s}^{(y)} {\rm \; are \; real} \:.
\end{equation}

\subsection{Magneto-dipole interaction}

The optical transitions between the spin branches can also occur due to the interaction of carriers with the magnetic field $\bm B$ of the incident electromagnetic wave. The Hamiltonian of the magneto-dipole interaction in the ($\psi_{0 +1/2}, \psi_{0 -1/2}$) basis is the Zeeman Hamiltonian
\begin{equation}
\label{eq:Ham_md}
\mathcal H_{\rm edge}^{(\bm B)} = - \bm{\mu} \cdot \bm{B}  = \frac{\mu_B}{2} \sum \limits_{\alpha, \beta = x,y,z} g_{\alpha \beta} \sigma_\alpha B_\beta\:,
\end{equation}
where $\bm{\mu}$ is the magnetic dipole operator, $g_{\alpha \beta}$ are the components of the $g$-factor tensor and $\mu_B$ is the Bohr magneton. Symmetry analysis (see Tab.~\ref{tab1}) shows that the non-zero components of the $g$-factor tensor are
$g_{xx}$, $g_{yy}$, $g_{zz}$, $g_{yz}$, and $g_{zy}$ in the structure of the $C_2$ group
and $g_{xx}$, $g_{yy}$, $g_{zz}$, $g_{xz}$, and $g_{zx}$ in the structure of the $C_s$ group. This is in agreement with
the results of microscopic calculations of Ref.~\onlinecite{Durnev2016}. For a structure with an arbitrarily oriented edge, all the components of the $g$-factor tensor can be non-zero.

\subsection{Microscopic description}\label{Sec_micdescription}

In this section we present the microscopic calculations of the matrix elements of the electric dipole operator $\bm{d}$ 
phenomenologically introduced in Eqs.~\eqref{d_C2} and~\eqref{d_Cs}. We consider TIs based on HgTe/CdHgTe QWs of 
the close-to-critical thickness. In such structures, the topological states are formed from the electron-like 
$|E1,\pm 1/2 \rangle$ and heavy-hole $|H1, \pm 3/2 \rangle$ subbands~\cite{Bernevig2006}.  In the basis 
$|E1,+ 1/2 \rangle$, $|H1, + 3/2 \rangle$, $|E1,- 1/2 \rangle$, and $|H1, - 3/2 \rangle$, the electron states 
in the QW of the D$_{2d}$ symmetry are described by the effective 4$\times$4
Hamiltonian, which takes into account the lack of the space inversion center in the QW~\cite{Durnev2016},
\begin{widetext}
\begin{equation}
\label{eq:H_bulk}
\mathcal H_0(k_x,k_y) =
\left( 
\begin{array}{cccc}
\delta_0 - (\B+\D)k^2 & {\rm i} \A k_+ & 0 & {\rm i} \gamma \e^{-2\i\theta} \\
-{\rm i} \A k_- & - \delta_0 + (\B-\D)k^2 & {\rm i} \gamma \e^{-2\i\theta} & 0\\
0 & -{\rm i} \gamma \e^{2\i\theta} & \delta_0 - (\B+\D)k^2 & - {\rm i} \A k_- \\
-{\rm i} \gamma \e^{2\i\theta} & 0 & {\rm i} \A k_+ & - \delta_0 + (\B-\D)k^2
\end{array}
\right) \: .
\end{equation}
\end{widetext}
Here, $\bm k = (k_x, k_y)$ is the in-plane electron wave vector, $k = |\bm k|$, $k_\pm = k_x \pm \i k_y$, $\A$, $\B$, $\D$, $\gamma$, and {$\delta_0$} are the real-valued band-structure parameters. In particular, the parameter $\delta_0$ describes the band gap and defines whether the system is in the trivial ($\delta_0 >0$ at $\B < 0$) or non-trivial ($\delta_0 < 0$ at $\B < 0$) topological phase~\cite{Bernevig2006}. The lack of the inversion center is taken into account by the parameter $\gamma$ which is microscopically determined by the strength of the mixing of the $|E1\rangle$ and $|H1\rangle$ states at the QW interfaces~\cite{Tarasenko2015}. To allow the consideration of the structures with
an arbitrary edge orientation, the Hamiltonian~\eqref{eq:H_bulk} is written in the coordinate frame ($xy$) rotated by the angle $\theta$
with respect to the crystallographic frame $([100], [010])$.

To calculate the wave functions of the edge states we consider a semi-infinite structure ($x \geq 0$) and solve the Schr\"odinger equation $\mathcal H_0 (- \i \partial/\partial x, k_y) \psi_{k_ys} = \eps_{k_ys} \psi_{k_ys}$ with the boundary conditions $\psi_{k_y s}(x=0,y) = 0$ and $\psi_{k_ys}(x \rightarrow + \infty,y) \rightarrow 0$.
The four-component wave functions $\psi_{k_ys}$ can be presented in the form
\begin{eqnarray}
\label{eq:wfs_general}
\psi_{k_y +1/2} &=& \frac{e^{\i k_y y}}{\sqrt{L}} \left(
\begin{array}{c}
a(x) \\
-b(x)  \\
-\i c(x) \e^{2 \i \theta} \\
-\i d(x) \e^{2 \i \theta}
\end{array}
\right) ,\:\:\: \nonumber\\
\psi_{k_y-1/2} &=& \frac{e^{\i k_y y}}{\sqrt{L}} \left(
\begin{array}{c}
-\i c(x) \e^{-2 \i \theta} \\
\i d(x) \e^{-2 \i \theta} \\
a(x)  \\
b(x)
\end{array}
\right) ,
\end{eqnarray}
where $a(x)$, $b(x)$, $c(x)$, and $d(x)$ are real functions, which also depend on $k_y^2$, and $L$ is the normalization length. 

The wave functions $\psi_{k_y +1/2}$ and $\psi_{k_y -1/2}$ given by Eqs.~\eqref{eq:wfs_general} are related to each other by the time reversal operator $\cal T$, see Eq.~\eqref{T-symmetry}, because ${\cal T}|E1, \pm 1/2\rangle = \mp |E1, \mp 1/2\rangle$ and 
${\cal T}|H1, \pm 3/2\rangle = \pm |H1, \mp 3/2\rangle$. Moreover, for high-symmetry edge directions, the functions $\psi_{k_y +1/2}$ and $\psi_{k_y -1/2}$ are transformed according to the rules introduced in Sec.~\ref{edge-symmetry}. In particular, for the structures
with the edge parallel to $\langle 010 \rangle$ ($\theta = \pi n/2$ with integer $n$), the wave functions are additionally related by the rotation operator $\cal R$, see Eq.~\eqref{R-rotation}, because ${\cal R}|E1, \pm 1/2\rangle = - \i |E1, \mp 1/2\rangle$ and 
${\cal R}|H1, \pm 3/2\rangle = \i |H1, \mp 3/2\rangle$. For the structures with the edge parallel to $\langle 110 \rangle$
($\theta = \pi/4 + \pi n/2$), the wave functions  are related by the reflection operator $\tilde{\cal R}$, 
see Eq.~\eqref{R-reflection}, because $\tilde{{\cal R}}|E1, \pm 1/2\rangle = \pm \i |E1, \mp 1/2\rangle$ and 
$\tilde{{\cal R}}|H1, \pm 3/2\rangle = \mp \i |H1, \mp 3/2\rangle$. 

The envelope functions $a(x)$, $b(x)$, $c(x)$, and $d(x)$ have to be calculated numerically. For the specific case of electron-hole symmetry, which corresponds to $\D = 0$ in the Hamiltonian~\eqref{eq:H_bulk}, the functions satisfy the relations $a(x) = b(x)$ and 
$c(x) = d(x)$.

The inter-branch matrix elements $v_{s -s}^{(x)}$ and $v_{s -s}^{(y)}$ of the velocity operator
\begin{equation}
\label{eq:vel_def}
{\bm v} = \frac{1}{\hbar} \frac{\partial \mathcal H_0}{\partial \bm k} \:
\end{equation}
can be directly calculated using the Hamiltonian~\eqref{eq:H_bulk} and the wave functions~\eqref{eq:wfs_general}. 
This procedure gives
\begin{eqnarray}\label{uxuy}
v_{s-s}^{(x)} = u_{1} \e^{-4is \theta} \:,\;\; v_{s-s}^{(y)} = 2\i s  u_{2} \, \e^{-4is \theta} \:,
\end{eqnarray}
where $u_{1}$ and $u_{2}$ are real quantities. 
Recalling the relation $\bm v_{s\, -s} = \i (\omega_{s \, -s}/e) \bm d_{s \, -s}$ we conclude that, within this microscopic model, the inter-branch optical transitions in the electric dipole approximation
are described by the dipole operator components which, at small $k_y$, read
\begin{eqnarray}\label{dxdy}
d_x &=&  (\sigma_y \cos 2\theta - \sigma_x \sin 2\theta) D_{1} k_y \:, \nonumber \\
d_y &=&  (\sigma_x \cos 2\theta + \sigma_y \sin 2\theta) D_{2} k_y \:, 
\end{eqnarray}
where $D_{1} = e u_{1} / | k_y \omega_{s \, -s}|$ and $D_{2} = e u_{2} / |k_y \omega_{s \, -s}|$ are independent of $k_y$.
Equations~\eqref{dxdy} are written for the edge of arbitrary orientation. At $\theta = \pi n/2$, they correspond to 
Eq.~\eqref{d_C2} with $D_{xy}=(-1)^n D_{1}$ and $D_{yx} = (-1)^n D_{2}$.
At $\theta = \pi/4 + \pi n/2$, they correspond to Eq.~\eqref{d_Cs} with $D_{xx}=(-1)^{n+1} D_{1}$ and $D_{yy} = (-1)^n D_{2}$.

The Zeeman Hamiltonian for the D$_{2d}$ QW in an in-plane magnetic field in the same basis
of the states as used for the Hamiltonian~\eqref{eq:H_bulk} has the form~\cite{Durnev2016} 
\begin{equation}
\label{eq:H_Binplane}
\mathcal H_{Z} = 
\frac{\mu_B}{2} \left(
\begin{array}{cccc}
0 & 0 & g_e^\parallel B_- & 0 \\
0 & 0 & 0 & g_h^\parallel \e^{-4 \i \theta} B_+ \\
g_e^\parallel B_+ & 0 & 0 & 0 \\
0 & g_h^\parallel \e^{4 \i \theta} B_- & 0 & 0
\end{array}
\right) \:,
\end{equation}
where $g_e^\parallel$ and $g_h^\parallel$ are 
the in-plane $g$-factors of the $|E1 \rangle$ and $|H1 \rangle$ subbands, which stem from
the bare electron $g$-factor and the interaction with 
remote electron and hole subbands, and $B_{\pm} = B_x \pm i B_y$. 
By projecting the Zeeman Hamiltonian~\eqref{eq:H_Binplane} onto the wave functions~\eqref{eq:wfs_general} we obtain the dependence 
of the edge-electron $g$-factor tensor on the edge orientation 

\begin{eqnarray}\label{g_all}
g_{xx} &=& g_{1} \cos^2 2\theta + g_{2} \sin^2 2\theta\:, \nonumber \\
g_{yy} &=& g_{1} \sin^2 2\theta + g_{2} \cos^2 2\theta\:, \nonumber \\
g_{xy} &=& g_{yx} = \frac12 \left( g_{1} - g_{2} \right) \sin 4 \theta\:,
\end{eqnarray}
where $g_{1}$ and $g_{2}$ are two independent $g$-factors. Analytical expressions for $g_{1}$ and $g_{2}$ via the band structure parameters are derived in Ref.~\onlinecite{Durnev2016} for the case of electron-hole symmetry, $\D = 0$, and $k_y = 0$.

\begin{figure}[htpb]
\includegraphics[width=0.45\textwidth]{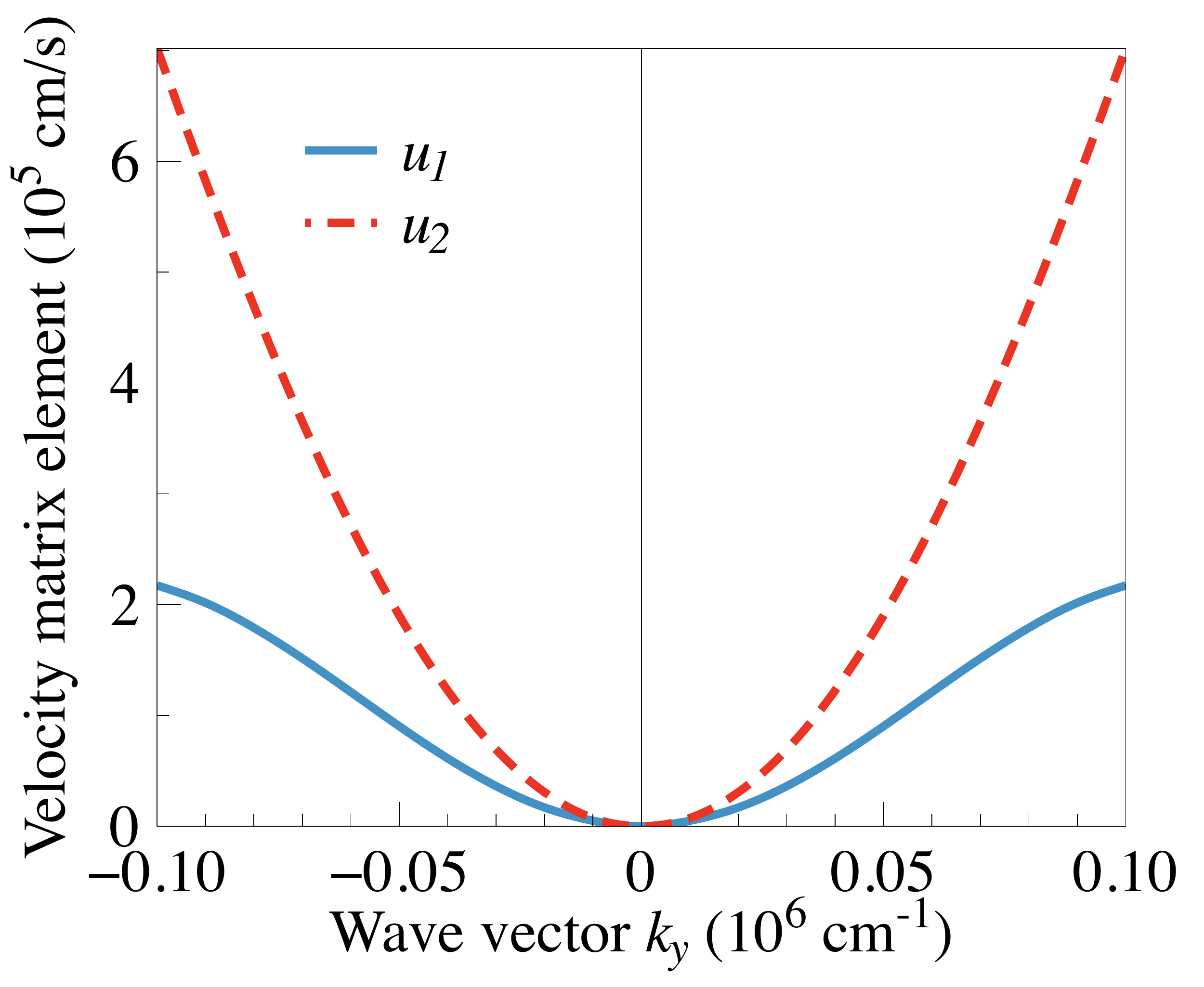}
\caption{\label{fig:fig3} Matrix elements of the velocity operator $u_1$ and $u_2$, see Eq.~\eqref{uxuy}, between the ``spin-up'' and 
``spin-down'' branches of the edge-state spectrum as a function of the electron wave vector along the edge $k_y$ calculated for the HgTe/CdHgTe QW structure.} 
\end{figure}

To evaluate the parameters $D_{1,2}$ and $g_{1,2}$ for realistic HgTe/CdHgTe QWs we numerically solve the Schr\"odinger equation with the
Hamiltonian~\eqref{eq:H_bulk}, calculate the wave functions of the edge states, the inter-branch matrix elements 
of the velocity operator, and the effective $g$-factors.  We use the following set of band-structure parameters: $\A = 3.6$~eV$\cdot$\AA, $\B = -68$~eV$\cdot$\AA$^2$, $\D = -51$~eV$\cdot$\AA$^2$~[\onlinecite{Konig2008}], $\gamma = 5$~meV~[\onlinecite{Tarasenko2015}], and $\delta = -10$~meV, which corresponds to the topological gap of about $20$~meV. Figure~\ref{fig:fig3} shows the calculated dependences of the inter-branch
matrix elements of the velocity operator $u_1$ and $u_2$, see Eq.~\eqref{uxuy}, on the wave vector $k_y$. At small $k_y$, the dependences are parabolic, which is in agreement with the phenomenological Eq.~\eqref{v_C2}. At large $k_y$, the dependences $u_1(k_y)$ and $u_2(k_y)$  deviate from the quadratic power due to high-order terms neglected in Eq.~\eqref{v_C2}. The fitting of the dependences at small $k_y$ yields $|D_1/e| \approx 7\times10^{-13}$~cm$^2$ and $|D_2 /e| \approx 1.4\times10^{-12}$~cm$^{2}$. Our numerical calculations also give $v_0 \approx 2.7\times10^7$~cm/s for the edge-state velocity and $|g_{1}| \approx 2.6$ and $|g_{2}| \approx 2$ for the effective $g$-factors.

\subsection{Particle-hole and chiral symmetries}

In addition to time reversal symmetry and the constraints imposed by spatial symmetry, the effective Hamiltonian may possess the particle-hole symmetry~\cite{Schnyder2008}. In this case, there is a charge conjugation operator $\mathcal C$ which anti-commutes with the Hamiltonian and can be presented in the form
$\mathcal C = U_c K$, where $U_c$ is a unitary operator
and $K$ is the operator of complex conjugation. The operator $\mathcal C$ relates the states $|\eps, k_y, s \rangle$ and $|-\eps, -k_y, s \rangle$
implying that $\eps_{k_y,s} = - \eps_{-k_y,s}$.

If both $\mathcal T$- and $\mathcal C$-symmetries are present, then the Hamiltonian also possesses the chiral (also called sublattice) symmetry.
The corresponding operator $\mathcal S = \mathcal C \mathcal T^{-1}$ anti-commutes with the Hamiltonian and satisfies $\mathcal S^\dag = \mathcal S^{-1}$ and $\mathcal S \mathcal S = 1$. The operator $\mathcal S$ relates the states $|\eps, k_y, s \rangle$ and $|-\eps, k_y, -s \rangle$. Its action on the edge state wave functions 
can be presented as 
\begin{equation}
\label{P-symmetry}
\mathcal S \psi_{\eps k_y s} = \e^{2 \i \alpha s} \psi_{-\eps k_y -s}\:,
\end{equation}
where $\alpha$ is a real parameter which may depend on $k_y$. If $\mathcal T$ and $\mathcal S$ are the only symmetries present in the system, $\alpha$ may be set to zero by a proper choice of the wave function phase. However, if there are additional spatial symmetries and the wave function phase is already fixed, see Eq.~\eqref{R-rotation} or Eq.~\eqref{R-reflection}, $\alpha$ is non-zero.

Equation~\eqref{P-symmetry} together with $\mathcal S \bm v = - \bm v \mathcal S$ implies that the inter-branch matrix elements of the velocity operator satisfy 
the relation
\begin{equation}
\label{v_P_props}
\bm v_{s-s}(k_y) = - \e^{-4\i \alpha s}  \bm v^*_{s -s}(k_y) \:,
\end{equation}
because $\bracket{\psi_{k_y s}}{\bm v}{\psi_{k_y -s}} = \e^{-4\i \alpha s} \bracket{\mathcal S \psi_{k_y -s}}{\bm v}{\mathcal S \psi_{k_y s}} =
- \e^{-4\i \alpha s} \bracket{\psi_{k_y -s}}{\bm v}{\psi_{k_y s}}$\:. It follows from Eq.~\eqref{v_P_props} that the ratio $v_{s -s}^{(x)}(k_y)/v_{s -s}^{(y)}(k_y)$ is a real value. 
Therefore, for circularly polarized radiation the probability of the optical transitions in the electro-dipole approximation is independent of the photon helicity sign and the circular (photon helicity dependent) photogalvanic effect is absent. Moreover, for linearly polarized radiation the absorption occurs only
for the radiation polarized along the in-plane direction $\bm n$ determined by $n_x/n_y = v_{s -s}^{(x)}(k_y)/v_{s -s}^{(y)}(k_y)$ while the orthogonally
polarized radiation is not absorbed. Due to time reversal symmetry, see Eq.~\eqref{Tsym}, the optical transitions induced by linearly polarized radiation
at the wave vectors $k_y$ and $-k_y$ have the same selection rules and the same probability.

The conclusions above are general and do not depend on the particular structure of edge states (as far as the boundary potential preserves $\mathcal T$- and $\mathcal{S}$-symmetries). In fact, they are also valid for systems of any dimension for the optical transitions between the states $|\eps, \bm k, s\rangle$ and $|-\eps, \bm k,-s\rangle$ related by chiral symmetry. Moreover, the same conclusions can be drawn for the optical transitions in the magneto-dipole approximation if the effective Zeeman Hamiltonian of the electron-photon interaction preserves $\mathcal S$-symmetry.

For the edges of particular crystallographic orientations, additional restrictions on the matrix elements of the velocity operator are imposed by spatial symmetry, see the relations~\eqref{vC2_props} and~\eqref{vCs_props} for the structures of the $C_2$ and $C_s$ point groups, respectively. Combining these relations with Eq.~\eqref{v_P_props} we conclude that,
in both cases, one of the matrix elements, either $v_{s-s}^{(x)}$ or $v_{s-s}^{(y)}$, vanishes.

The effective Hamiltonian~\eqref{eq:H_bulk} possesses the particle-hole symmetry at $\D = 0$. The corresponding unitary matrix $U_c$ of the charge conjugation operator $\mathcal C$ reads
\begin{equation}
\label{eq:Umatrix}
U_c = \left(
\begin{array}{cccc}
0 & \e^{-2\i\theta} & 0 & 0 \\
\e^{-2\i\theta} & 0 & 0 & 0 \\
0 & 0 & 0 & \e^{2\i\theta}\\
0 & 0 & \e^{2\i\theta} & 0
\end{array}
\right)\:.
\end{equation}
Since $U_c^T = U_c$ and hence $\mathcal C \mathcal C = 1$, the system under study belongs to the DIII symmetry 
class~\cite{Schnyder2008}.

The matrix $\mathcal S = U_c U_t^{-1}$, which satisfies $\mathcal S \mathcal H_0 + \mathcal H_0 \mathcal S = 0$ at
$\D = 0$, has the form
\begin{equation}
\label{eq:Smatrix}
\mathcal S = \left(
\begin{array}{cccc}
0 & 0 & 0 & \e^{-2\i\theta} \\
0 & 0 & -\e^{-2\i\theta} & 0 \\
0 & -\e^{2\i\theta} & 0 & 0\\
\e^{2\i\theta} & 0 & 0 & 0
\end{array}
\right)\:.
\end{equation}
It relates the wave functions~\eqref{eq:wfs_general} by $\mathcal S \psi_{k_y s} = \e^{4 \i \theta s} \psi_{k_y -s}$, so that $\alpha = 2\theta$ in Eqs.~\eqref{P-symmetry} and~\eqref{v_P_props}. Since $v_{s-s}^{(x)} = u_x \e^{-4is \theta}$ and $v_{s-s}^{(y)} = 2\i s u_y \e^{-4is \theta}$, it follows from Eq.~\eqref{v_P_props} that $v_{s-s}^{(x)} = 0$ for any orientation of the edge. Hence, in the framework of the effective Hamiltonian~\eqref{eq:H_bulk}, the electro-dipole transitions for the radiation polarized perpendicular to the edge are forbidden at $\D = 0$.

In real semiconductor structures, particle-hole symmetry is broken and both matrix elements $v_{s-s}^{(x)}$ and $v_{s-s}^{(y)}$ are non-zero. Therefore, the optical transitions between the spin branches of the helical channel are allowed for radiation polarized along the edge and perpendicular to the edge. Moreover, the phase shift of $\pm \pi/2$ between $v_{s-s}^{(x)}$ and $v_{s-s}^{(y)}$, see Eq.~\eqref{uxuy} [or between the matrix elements of the electric dipole components $d_x$ and $d_y$, see Eq.~\eqref{dxdy}], means that the transitions are sensitive to the degree and the sign of circular polarization.

\section{Optical transitions. Linear and circular dichroisms}\label{Sec3}

Consider now that the topological insulator is illuminated by radiation of certain frequency and polarization
which induces direct optical transitions between the spin branches of the helical edge channel.
With the account for both the electro-dipole and the magneto-dipole mechanisms of electron-photon interaction,
the optical transitions $|k_y, -s\rangle \rightarrow |k_y, s\rangle$ are described by the matrix elements
\begin{equation}
M_{s\, -s} (k_y)= - \bm{d}_{s\, -s} \cdot \bm{E}_0 - \bm{\mu}_{s\, -s} \cdot \bm{B}_0 \,,
\end{equation}
where $\bm{d}_{s\, -s}$ and $\bm{\mu}_{s\, -s}$ are the matrix elements of the electric dipole and magnetic dipole operators,
respectively, $\bm{E}_0$ and $\bm{B}_0$ are the amplitudes of the electric and magnetic fields of the radiation related by
$\bm{B}_0 = n_{\omega} \, \bm{o} \times \bm{E}_0$, $n_{\omega}$ is the refractive index of the medium, and $\bm{o}$ is the unit vector 
along the radiation propagation direction $\pm z$. 

It follows from Eqs.~\eqref{dxdy} and~\eqref{g_all} that, at small $k_y$, the matrix elements are given by 
\begin{eqnarray}
M_{\pm 1/2 \, \mp 1/2}(k_y) = \pm \i \e^{\mp 2\i \theta} (D_1 E_{0x} \pm \i D_2 E_{0y}) k_y \hspace{1.2cm} \\
+ \frac{\mu_B}{2} \left[ \frac{g_1 + g_2}{2} (B_{0x} \mp \i B_{0y}) + \frac{g_1 - g_2}{2} \e^{\mp 4 \i \theta}(B_{0x} \pm \i B_{0y}) \right] \:. \nonumber
\end{eqnarray}

The values $|M_{+1/2 \, - 1/2}(k_y)|^2$ and $|M_{-1/2 \, + 1/2}(-k_y)|^2$, which determine the probabilities of the optical transitions at $k_y$ and $-k_y$,
respectively, see Fig.~\ref{fig:fig2}, can be decomposed into the symmetric and asymmetric parts as follows
\begin{equation}
|M|_{\rm sym/asym}^2 = \frac{|M_{+1/2 \, - 1/2}(k_y)|^2 \pm |M_{-1/2 \, + 1/2}(-k_y)|^2}{2} .
\end{equation}
Straightforward calculations give
\begin{eqnarray}\label{M2_sym}
|M|_{\rm sym}^2 &=& ( D_1^2 |e_x|^2 + D_2^2 |e_y|^2 ) E_0^2 k_y^2 \\
&-& \frac{D_1 g_1 - D_2 g_2}{2} |k_y| \mu_B n_{\omega} \cos 2\theta \, E_0^2 P_{circ} \,, \nonumber
\end{eqnarray}
\begin{eqnarray}\label{M2_asym}
|M|_{\rm asym}^2 &=& - D_1 D_2 E_0^2 k_y^2 \, {\rm sign}k_y \, P_{circ} o_z  \\
&-& \frac{D_1 g_2 - D_2 g_1}{2} k_y \mu_B n_{\omega} \cos 2\theta \, E_0^2 o_z \nonumber  \\
&-& \frac{D_1 g_2 + D_2 g_1}{2}k_y \mu_B n_{\omega} \cos 2\theta \, E_0^2 (|e_x|^2 - |e_y|^2) o_z \nonumber \\
&-& \frac{D_1 g_2 + D_2 g_1}{2} k_y \mu_B n_{\omega} \sin 2\theta \, E_0^2 (e_x e_y^* + e_y e_x^* ) o_z \nonumber \,,
\end{eqnarray}
where $\bm{e} = \bm{E}_0/E_0$ is the (complex) unit vector of the radiation polarization and $P_{circ} = \i (e_x e_y^* - e_y e_x^*) o_z$ is the radiation
helicity. In Eqs.~\eqref{M2_sym} and~\eqref{M2_asym} we keep the terms originating from the electro-dipole interaction and from the interference of the electro-dipole and magneto-dipole interactions. The terms stemming solely from the magneto-dipole interaction are small and neglected. 

The absorption width of the edge channel is defined by
\begin{equation}
w = W/ I \,,
\end{equation}
where $W$ is the energy absorbed per unit time per unit length of the edge channel,
\begin{eqnarray}
W &=& 4\pi \omega \sum_{k_y > 0} |M|_{\rm sym}^2 [f(\eps_{k_y, -1/2})-f(\eps_{k_y, +1/2})] \nonumber \\
&\times& \delta(\eps_{k_y +1/2} - \eps_{k_y -1/2} - \hbar \omega) \,,
\end{eqnarray}
$f(\eps)$ is the Fermi-Dirac distribution function, and  $I = c n_{\omega} E_0^2 /(2\pi)$ is the radiation intensity.

The calculation of the absorption width for the probability of the optical transtions given by Eq.~\eqref{M2_sym} and the linear
dispersion $\eps_{k_y \pm1/2} = \pm \hbar v_0 k_y$ yields
\begin{eqnarray}
w &=& \frac{\pi \omega^3 \Delta f }{2c n_{\omega} \hbar v_0^3} \left( D_1^2 |e_x|^2 + D_2^2 |e_y|^2 \right) \\
&-& \frac{\pi \mu_B \omega^2 \Delta f }{2c \hbar v_0^2} (D_1 g_1 - D_2 g_2)  \cos 2\theta P_{circ}  \,, \nonumber
\end{eqnarray}
where $\Delta f = f(-\hbar\omega/2) - f(\hbar\omega/2)$.

We conclude that the edge of a topological insulator based on a zinc-blend-type crystal exhibits linear and circular dichroisms while the bulk material does not. 
For linearly polarized radiation, the absorption depends on the orientation of the polarization vector $\bm e$. The ratio of the absorption widths
for the radiation polarized along the edge ($\bm e \parallel y$) and perpendicular to the edge ($\bm e \parallel x$) is given by $(D_2/D_1)^2$ 
which is estimated as $4$ for HgTe/CdHgTe-based structures. For circularly polarized radiation, the absorption contains a contribution sensitive to the photon helicity $P_{circ}$, i.e., the absorption is different for right-handed and left-handed circularly polarized photons. Here, the effect stems from the interference of the electro-dipole and magneto-dipole mechanisms of the photon absorption. In accordance with the general theory of the circular dichroism, the effect is absent
in systems with mirror planes, which is realized in our structure if $\theta = \pi/4 + \pi n/2$. Interestingly, for a structure with any other $\theta$,
the circular dichroism occurs and has the same sign for the opposite edges of the structure.

\section{Spin polarization and Edge photocurrents}\label{Sec4}

The illumination of a topological insulator leads also to a spin polarization of electrons and a direct electric current in the edge channel 
since the optical transitions $|-k_y, +1/2\rangle \rightarrow |-k_y, -1/2\rangle$ and $|k_y, -1/2\rangle \rightarrow |k_y, +1/2\rangle$ occur at different rates
that is described by $|M|_{\rm asym}^2$, see Eq.~\eqref{M2_asym}. The mechanism of the current generation is illustrated in Fig.~\ref{fig:fig2}.

In the relaxation time approximation, the photocurrent is given by~\cite{Ganichev2003}
\begin{eqnarray}\label{eq:jy}
j_y &=& \frac{4 \pi e}{\hbar} \sum_{k_y > 0} [\tau_p (\eps_{k_y +1/2}) v_{k_y +1/2} - \tau_p (\eps_{k_y -1/2}) v_{k_y -1/2} ] \nonumber \\
&\times& |M|_{\rm asym}^2 [f(\eps_{k_y, -1/2})-f(\eps_{k_y, +1/2})] \nonumber \\
&\times& \delta(\eps_{k_y +1/2} - \eps_{k_y -1/2} - \hbar \omega) \,,
\end{eqnarray}
where $v_{k_y s} = (1/\hbar) d \eps_{k_y s} / d k_y$ is the intra-branch velocity and $\tau_p$ is the relaxation time of electrons in the edge channel which is determined by spin-flip processes. It is known that in real HgTe/CdHgTe structures, the topological protection against spin-flip scattering is violated and ballistic transport is observed only in $\mu$m-scale devices~\cite{Roth2009,Dantscher2017}. 

Taking into account that $\eps_{k_y \pm 1/2} = \pm \hbar v_0 k_y$ at small $k_y$ and calculating the sum over $k_y$ in Eq.~\eqref{eq:jy} we obtain
\begin{equation}\label{eq:j2}
j_y = \frac{2e \bar{\tau}_p}{\hbar^2} |M(\omega/2v_0)|_{\rm asym}^2 \Delta f \,,
\end{equation}
where $\bar{\tau}_p = [\tau_p(\hbar\omega/2) + \tau_p(-\hbar\omega/2)]$.

The photocurrent sensitive to the photon helicity emerges in the electro-dipole approximation. The substitution of the first line in Eq.~\eqref{M2_asym}
for $|M|_{\rm asym}^2$ in Eq.~\eqref{eq:j2} yields
\begin{equation}\label{j_circ_final}
j_y^{({\rm circ})} = - \frac{4e \bar{\tau}_p v_0 w_0}{\hbar \omega} \frac{D_1 D_2}{D_1^2 + D_2^2} I P_{circ} o_z \:,
\end{equation}
where $w_0$ is the absorption width of the edge channel for circularly polarized radiation calculated
in the electro-dipole approximation,
\begin{equation}
w_0 = \frac{\pi \omega^3 (D_1^2 + D_2^2) \Delta f}{4 c n_{\omega} \hbar v_0^3} \:.
\end{equation}
Equation~\eqref{j_circ_final} describes the circular photogalvanic effect~\cite{Ganichev2003,Wittmann2010,Ivchenko_book,SturmanFridkin} in helical edge channels. 
The photocurrent is proportional to the degree of circular polarization and is reversed by switching the sign of the photon helicity. We note, that the magneto-dipole transitions also give a contribution to the circular photocurrent~\cite{Dora2012, Artemenko2013}. However, this contribution is a few orders of magnitude smaller than the contribution of the electro-dipole transitions.

For linearly polarized radiation, the spin polarization of electrons and the corresponding electric current
emerge due to the interference of the electro-dipole and magneto-dipole transitions, see the second, third, and forth lines in Eq.~\eqref{M2_asym}.
This photocurrent has the form
\begin{eqnarray}\label{j_lin_final}
j_y^{({\rm lin})} =  \left[A + B \left( \left| e_x \right|^2 - \left| e_y \right|^2 \right)\right] \cos 2\theta \, I o_z \nonumber \\ 
+ B ( e_x e_y^* + e_y e_x^* ) \sin 2\theta \,  I o_z \:,
\end{eqnarray}
where
\begin{eqnarray}\label{eq:Q}
A &=& - \frac{4e \bar{\tau}_p v_0^2 \, w_0 n_{\omega}}{\hbar \omega^2} \frac{\mu_B (D_1 g_2 - D_2 g_1)}{D_1^2 + D_2^2} \:, \nonumber \\
B &=& -  \frac{4e \bar{\tau}_p v_0^2 \, w_0 n_{\omega}}{\hbar \omega^2} \frac{\mu_B (D_1 g_2 + D_2 g_1)}{D_1^2 + D_2^2}  \:. 
\end{eqnarray}
The linear photocurrent depends on the orientation of the edge with respect to the crystallographic axes and the radiation polarization plane with respect to the edge. It may also
appear when the sample is excited by unpolarized radiation. The linear photocurrent~\eqref{j_lin_final} originates from the action of both
ac electric and magnetic fields of the radiation upon electrons and belongs to the class of ac Hall effects~\cite{Barlow1954,Perel1973,Karch2010} or, more generally, to the class of photoelectric effects caused by light pressure (photon drag). In our case,
the photocurrent flows in a direction perpendicular to the photon wave vector. 

\begin{figure}[t]
\includegraphics[width=0.45\textwidth]{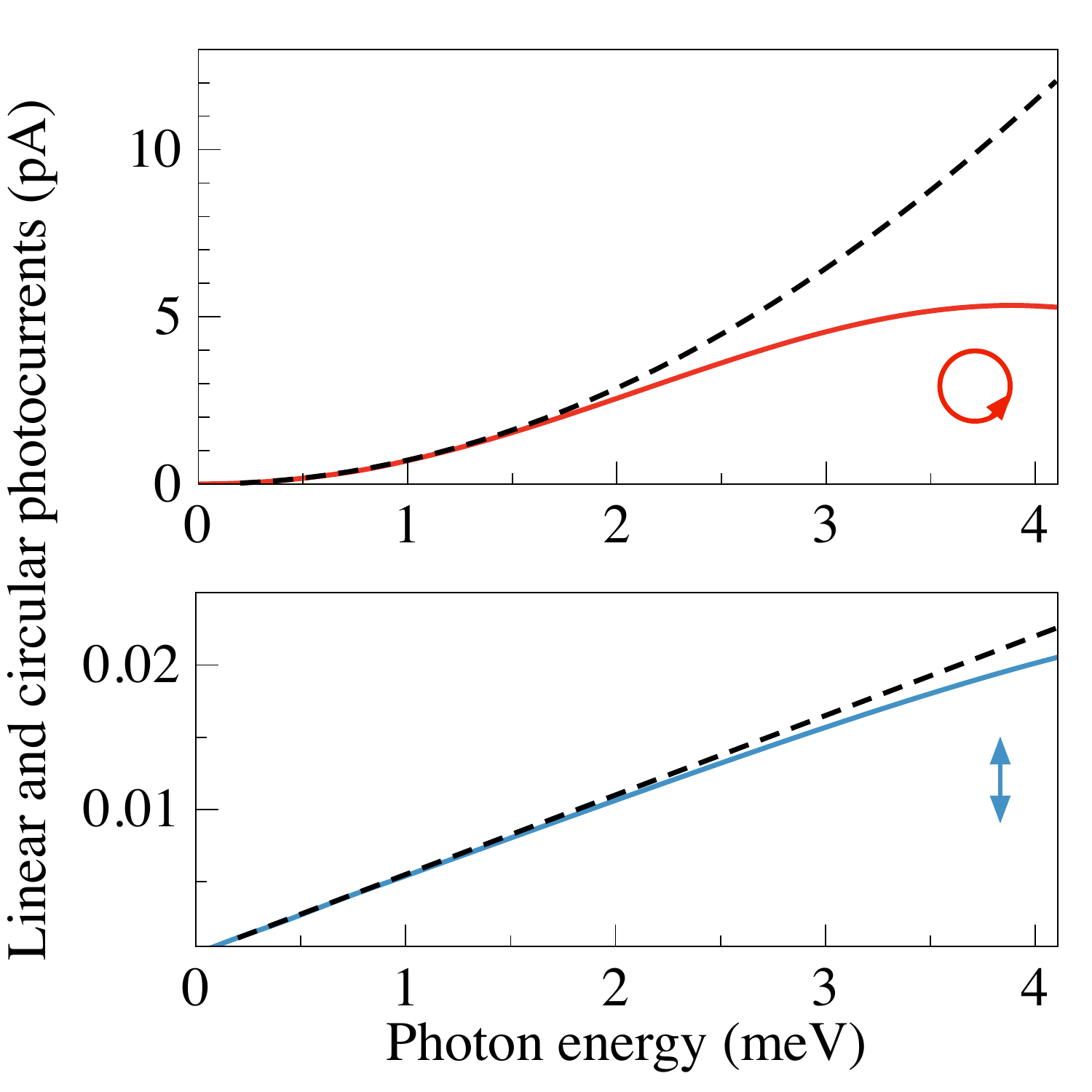}
\caption{\label{fig:fig4} 
Amplitudes of the circular and linear edge photocurrents as a function of the incident photon energy. The dependences are
calculated for the parameters of HgTe/CdHgTe-based two-dimensional topological insulators presented in Sec.~\ref{Sec_micdescription}, the momentum relaxation time
$\bar{\tau}_p = 20$~ps, the refractive index $n_{\omega} = 3$, and the radiation intensity $I=1$~W/cm$^2$. The linear photocurrent is calculated for $\theta = 0$
and $\bm e \parallel y$. Solid curves show the results of numerical calculations, dashed curves are plotted after analytical Eqs.~\eqref{j_circ_final} and~\eqref{j_lin_final}.
}
\end{figure}

Figure~\ref{fig:fig4} shows the amplitudes of the circular and linear edge photocurrents in a HgTe/CdHgTe topological insulator as a function of the photon energy 
$\hbar\omega$. The dependences are calculated for zero temperature, the Fermi level lying at the Dirac point, and the relaxation time $\bar{\tau}_p = 20$~ps estimated from the experiments~\cite{Dantscher2017}. Solid curves present the results based on numerical calculations of the matrix elements of the electron-photon interaction.
Dashed curves show the low-energy analytical results plotted after Eqs.~\eqref{j_circ_final} and~\eqref{j_lin_final}. For the radiation intensity 1~W/cm$^2$, the photon energy $2$~meV and the momentum relaxation time presented above, we expect the circular photogalvanic current of a few pA and the linear photon drag current 
of a ten fA.       

\section{Summary}

To summarize, we have theoretically investigated optical properties of helical edge channels in two-dimensional topological insulators based on zinc-blende quantum wells and edge photocurrents, which occur under illumination of edge channels. We have shown that the lack of space inversion symmetry in the zinc-blende-type quantum wells results in the electro-dipole optical transitions between the ``spin-up'' and ``spin-down'' branches of the helical channels. Using general symmetry arguments we have analyzed polarization dependence of these transitions for the structures with different edge orientations. Based on the extended Bernevig-Hughes-Zhang Hamiltonian we have performed the microscopic calculations of the inter-branch electric dipole matrix elements for HgTe/CdHgTe topological insulators. It has been shown that the asymmetry of electro-dipole optical transitions in $k$-space results in generation of circular edge photogalvanic current, which direction is controlled by the helicity of the incident photons. The photocurrent amplitude of this current is a few orders of magnitude larger than of the one expected from the magneto-dipole mechanism of optical transitions. From the general symmetry analysis we have shown that the circular edge photocurrent is absent in the structures, where the electron spectrum possesses particle-hole symmetry. It is also established that the interference of the electro-dipole and magneto-dipole mechanisms of inter-branch optical transitions results, under illumination with linearly polarized radiation, in the generation of linear edge photocurrent. Its direction and amplitude are controlled by the crystallographic orientation of the edge and the orientation of the polarization vector with respect to the edge. The interference of the electro-dipole and magneto-dipole optical transitions gives rise to the circular dichroism of the helical edge channels.
 
\acknowledgements

This work was supported by the Russian Science Foundation (project 17-12-01265). M.V.D. also acknowledges the financial support from the RFBR (project No.16-32-60175).



\begin{thebibliography}{99}

\bibitem{Bernevig2006}
B. A. Bernevig, T. L. Hughes, and S.-C. Zhang,  Quantum spin Hall Effect and topological phase transition in HgTe quantum wells, Science \textbf{314}, 1757 (2006).

\bibitem{Konig2007}
M. K\"onig, S. Wiedmann, C. Br\"une, A. Roth, H. Buhmann, L. W. Molenkamp, X.-L. Qi, and S.-C. Zhang,  Quantum spin Hall insulator state in HgTe quantum wells, Science \textbf{318}, 766 (2007).

\bibitem{Knez2011}
I. Knez, R.-R. Du, and G. Sullivan, Evidence for helical edge modes in inverted InAs/GaSb quantum wells, Phys. Rev. Lett. \textbf{107}, 136603 (2011).

\bibitem{Roth2009}
A. Roth, C. Br\"{u}ne, H. Buhmann, L. W. Molenkamp, J. Maciejko, X.-L. Qi, and S.-C. Zhang, 
Nonlocal transport in the quantum spin Hall state, Science \textbf{325}, 294 (2009).

\bibitem{Gusev2011}
G. M. Gusev, Z. D. Kvon, O. A. Shegai, N. N. Mikhailov, S. A. Dvoretsky, and J. C. Portal,
Transport in disordered two-dimensional topological insulators,
Phys. Rev. B \textbf{84}, 121302(R) (2011).

\bibitem{Ma2015} E. Y. Ma, M. R. Calvo, J. Wang, B. Lian, M. M\"{u}hlbauer, C. Br\"{u}ne, Y.-T. Cui, K. Lai, 
W. Kundhikanjana, Y. Yang, M. Baenninger, M. K\"{o}nig, C. Ames, H. Buhmann, P. Leubner, L. W. Molenkamp, S.-C. Zhang, 
D. Goldhaber-Gordon, M. A. Kelly, and Z.-X. Shen,
Unexpected edge conduction in mercury telluride quantum wells under broken time-reversal symmetry,
Nat. Comm. {\bf 6}, 7252 (2015).

\bibitem{Tikhonov2015} E. S. Tikhonov, D. V. Shovkun, V. S. Khrapai, Z. D. Kvon, N. N. Mikhailov, and
S. A. Dvoretsky, Shot noise of the edge transport in the inverted band HgTe quantum wells,
JETP Lett. \textbf{101}, 708 (2015).

\bibitem{Hart2014}
S. Hart, H. Ren, T. Wagner, P. Leubner, M. M\"uhlbauer, C. Br\"une, H. Buhmann, L. W. Molenkamp, and A. Yacoby,
Induced superconductivity in the quantum spin Hall edge, Nat. Phys. \textbf{10}, 638 (2014).

\bibitem{Kononov2015}
A. Kononov, S. V. Egorov, Z. D. Kvon, N. N. Mikhailov, S. A. Dvoretsky, and E. V. Deviatov,  Evidence on the macroscopic length scale spin coherence for the edge currents in a narrow HgTe quantum well, JETP Letters \textbf{101}, 814 (2015).

\bibitem{Tanaka2011} Y. Tanaka, A. Furusaki, and K. A. Matveev, 
Conductance of a helical edge liquid coupled to a magnetic impurity,
Phys. Rev. Lett. {\bf 106}, 236402 (2011).

\bibitem{Lunde2012} A. M. Lunde and G. Platero, Helical edge states coupled to a spin bath: Current-induced magnetization,
Phys. Rev. B {\bf 86}, 035112 (2012).

\bibitem{Altshuler2013} B.L. Altshuler, I.L. Aleiner, and V.I. Yudson, 
Localization at the edge of a 2D topological insulator by Kondo impurities with random anisotropies,
Phys. Rev. Lett. {\bf 111}, 086401 (2013). 

\bibitem{Vayrynen2014} J. I. V\"{a}yrynen, M. Goldstein, Yu. Gefen, and L. I. Glazman,
Resistance of helical edges formed in a semiconductor heterostructure,
Phys. Rev. B {\bf 90}, 115309 (2014).

\bibitem{Entin2015} M. V. Entin and L. I. Magarill,
Localization of edge electrons in a 2D topological insulator strip,
JETP Lett. {\bf 100},  566 (2015).

\bibitem{Kurilovich2017} P. D. Kurilovich, V. D. Kurilovich, I. S. Burmistrov, and M. Goldstein,
Helical edge transport in the presence of a magnetic impurity,
JETP Lett. {\bf 106}, 593 (2017).

\bibitem{Dantscher2017}
K.-M. Dantscher, D. A. Kozlov, M. T. Scherr, S. Gebert, J. B\"arenf\"anger, M. V. Durnev, S. A. Tarasenko, V. V. Bel'kov, N. N. Mikhailov, S. A. Dvoretsky, Z. D. Kvon, J. Ziegler, D. Weiss, and S. D. Ganichev.  Photogalvanic probing of helical edge channels in two-dimensional HgTe topological insulators, Phys. Rev. B \textbf{95}, 201103 (2017).

\bibitem{Dora2012}
B. D\'ora, J. Cayssol, F. Simon, and R. Moessner.  Optically Engineering the Topological Properties of a Spin Hall Insulator, Phys. Rev. Lett. \textbf{108}, 056602 (2012).

\bibitem{Artemenko2013}
S. N. Artemenko and V. O. Kaladzhyan. Photogalvanic effects in topological insulators, JETP Letters \textbf{97}, 82 (2013).

\bibitem{Tarasenko2015}
S. A. Tarasenko, M. V. Durnev, M. O. Nestoklon, E. L. Ivchenko, J.-W. Luo, and A. Zunger,  Split Dirac cones in HgTe/CdTe quantum wells due to symmetry-enforced level anticrossing at interfaces, Phys. Rev. B \textbf{91}, 081302 (2015).

\bibitem{Durnev2016}
M. V. Durnev and S. A. Tarasenko.  Magnetic field effects on edge and bulk states in topological insulators based on HgTe/CdHgTe quantum wells with strong natural interface inversion asymmetry, Phys. Rev. B \textbf{93}, 075434 (2016).

\bibitem{koster63}
G. F. Koster, R. G. Wheeler, J. O. Dimmock, and H. Statz. \textit{Properties of Thirty-Two Point Groups} (MIT Press, Cambridge, MA, 1963).

\bibitem{birpikus}
G. L. Bir and G. E. Pikus. \textit{Symmetry and Strain-Induced Effects in Semiconductors}  (Wiley, New York, 1974).

\bibitem{Teppe2018}
S. S. Krishtopenko and F. Teppe.  Realistic picture of helical edge states in HgTe quantum wells, Phys. Rev. B \textbf{97}, 165408 (2018).

\bibitem{Konig2008}
M. K\"onig, H. Buhmann, L. W. Molenkamp, T. Hughes, C.-X. Liu, X.-L. Qi, and S.-C. Zhang, The quantum spin Hall effect: Theory and experiment, J. Phys. Soc. Jpn. \textbf{77}, 031007 (2008).

\bibitem{Schnyder2008}
A. P. Schnyder, S. Ryu, A. Furusaki, and A. W. W. Ludwig.  Classification of topological insulators and superconductors in three spatial dimensions, Phys. Rev. B \textbf{78}, 195125 (2008).

\bibitem{Ganichev2003} S. D. Ganichev, V. V. Bel'kov, P. Schneider, E. L. Ivchenko, S. A. Tarasenko, W. Wegscheider, D. Weiss, 
D. Schuh, E. V. Beregulin, and W. Prettl, 
Resonant inversion of the circular photogalvanic effect in n-doped quantum wells, 
Phys. Rev. B {\bf 68}, 035319 (2003).

\bibitem{Wittmann2010} B. Wittmann, S. N. Danilov, V. V. Bel'kov, S. A. Tarasenko, E. G. Novik, H. Buhmann, C. Br\"{u}ne, L. W. Molenkamp, 
Z. D. Kvon, N. N. Mikhailov, S. A. Dvoretsky, N. Q. Vinh,  A. F. G. van der Meer, B. Murdin, and S. D. Ganichev, 
Circular photogalvanic effect in HgTe/CdHgTe quantum well structures, 
Semicond. Sci. Technol. {\bf 25}, 095005 (2010).

\bibitem{Ivchenko_book} E. L. Ivchenko, {\it  Optical Spectroscopy of Semiconductor Nanostructures} (Harrow, UK: Alpha Science Int., 2005).

\bibitem{SturmanFridkin} B. I. Sturman and V. M. Fridkin, {\it The Photovoltaic and Photorefractive Effects in Non-Centrosymmetric Materials}
(NY: Gordon and Breach, 1992).

\bibitem{Barlow1954} H. M. Barlow,
Application of the Hall effect in a semiconductor to the measurement of power in an electromagnetic field,
Nature (London) {\bf 173}, 41 (1954).

\bibitem{Perel1973} V. I. Perel' and Ya. M. Pinskii, 
Constant current in conducting media due to a high-frequency electromagnetic field,
Sov. Phys. Solid State {\bf 15}, 688 (1973).

\bibitem{Karch2010} J. Karch, J. Karch, P. Olbrich, M. Schmalzbauer, C. Zoth, C. Brinsteiner, M. Fehrenbacher, U. Wurstbauer, M.M. Glazov, S.A. Tarasenko, E.L. Ivchenko, D. Weiss, J. Eroms, R. Yakimova, S. Lara-Avila, S. Kubatkin, and S.D. Ganichev, 
Dynamic Hall effect driven by circularly polarized light in a graphene layer, 
Phys. Rev. Lett. {\bf 105}, 227402 (2010).






\end{thebibliography}
\end{document}